\documentclass[sn-mathphys]{sn-jnl}
\jyear{2021}%
\usepackage{mathptmx}
\DeclareUnicodeCharacter{2212}{-}
\usepackage{amsmath}
\usepackage{bm}

\newcounter{Mequ}
\newenvironment{MEquation}
  {\stepcounter{Mequ}%
    \addtocounter{equation}{-1}%
    \renewcommand\theequation{M\arabic{Mequ}}\equation}
  {\endequation}
\begin{document}

\title[Bayesian Testing Of Granger Causality In Functional Time Series]{Bayesian Testing Of Granger Causality In Functional Time Series}
\author*[1]{\fnm{Rituparna} \sur{Sen}}\email{ritupar.sen@gmail.com}

\author[2]{\fnm{Anandamayee} \sur{Majumdar}}\email{amajumdar@iattc.org}

\author[3]{\fnm{Shubhangi} \sur{Sikaria}}\email{shubhangisikariya@gmail.com}

\affil*[1]{\orgdiv{Applied Statistics Unit}, \orgname{Indian Statistical Institute}, \orgaddress{\street{ 8th Mile, Mysore Rd, RVCE Post}, \city{Bangalore}, \postcode{560059}, \state{KA}, \country{India}}}

\affil[2]{\orgdiv{Stock Assessment Program}, \orgname{Inter-American Tropical Tuna Commission}, \orgaddress{\city{San Diego}, \state{CA}, \country{USA}}}

\affil[3]{\orgdiv{Department of Mathematics}, \orgname{Indian Institute of Technology Madras}, \orgaddress{\city{Chennai}, \state{TN}, \country{India}}}

\abstract{We develop a multivariate functional autoregressive model (MFAR), which captures the cross-correlation among multiple functional time series and thus improves forecast accuracy. We estimate the parameters under the Bayesian dynamic linear models (DLM)  framework. In order to capture Granger causality from one FAR series to another we employ Bayes Factor. Motivated by the broad application of functional data in finance, we investigate the causality between the yield curves of two countries. Furthermore, we illustrate a climatology example, examining whether the weather conditions Granger cause pollutant daily levels in a city.}

\keywords{Multivariate functional time series, dynamic linear model, Granger causality, Bayesian Analysis}

\maketitle
\section{Introduction}
A functional time series arises when at each point in time we observe a random function, generally at discrete points on a dense support. Examples of functions include satellite images(Guyet \& Nicolas \cite{guyet2016long}), fMRI data(Stoehr, Aston \& Kirch \cite{stoehr2021detecting}) and yield curves( Sen \& Kl{\"u}ppelberg \cite{sen2019time}). Another common practice that leads to functional time series is to slice high-frequency temporal sequence records into curves over logical consecutive time intervals. Examples of such curve time series include the daily price curve of financial transactions ( Horv{\'a}th, Kokoszka \& Rice \cite{horvath2014testing}; Li, Robinson \& Shang \cite{li2020long}), electricity price curves as a function of day-of-year ( Chen \& Li \cite{chen2017adaptive}; Chen, Marron \& Zhang \cite{chen2019modeling}), daily pollution levels as a function of the time of the day (Aue, Norinho \& H{\"o}rmann \cite{aue2015prediction}), and daily patterns of environmental data (Besse, Cardot \& Stephenson \cite{besse2000autoregressive}, Ramsay \& Dalzell \cite{ramsay1991some}). The functional autoregressive (FAR) model, developed by Bosq \cite{bosq2000linear}, plays a central role in modeling and predicting functional time series. The FAR model extends the vector autoregressive (VAR) model to the infinite-dimensional space and considers the serial correlation between the functional observations.

Most of the existing literature relies on functional principal component analysis (FPCA) to treat multivariate functional data. In a multivariate setting, along with serial correlation, one needs to study the correlation among the data series. Ramsay \& Silverman \cite{ramsay2005principal} introduced a classical multivariate FPCA that concatenates the multiple functional data and performs univariate FPCA on a single long vector. However, this model is restricted to multivariate random functions on the same finite, one-dimensional interval. Chiou, Chen \& Yang \cite{chiou2014multivariate} discussed a normalized multivariate functional principal component analysis (mFPCA$_n$) which accounts for data measured in different units using normalized covariance operators. Although mFPCA$_n$ produces sound estimates on dense grids without calculation errors, it leads to inadmissible results with significant measurement errors in the case of sparse settings.

Apart from FPCA, there is abundant literature that models multivariate functional data using regression analysis. Zhu, Li \& Kong \cite{zhu2012multivariate} developed a multivariate varying coefficients model (MVCM) to estimate covariance function. They also establish the uniform convergence rate of this covariance function. Li et al. \cite{li2017functional} extended the previous varying-coefficient single index model to perform the regression on functional response data. Zhu et al. \cite{zhu2017multivariate} performed linear transformation and adopted the Bayesian approach to sample the model parameters in the transformed space. Zhu, Strawn, \& Dunson \cite{zhu2016bayesian} introduced a notion of graphical models in multivariate functional data by constructing Bayesian inference of undirected, decomposable graphs based on Markov distribution. The literature mentioned above mainly focuses on dense functional data. Li, Xiao \& Luo \cite{li2020fast} adopted the fast covariance-based FPCA method for multivariate sparse functional data, approximating covariance functions using Tensor-product B-splines. The underlying assumption while using FPCA is that the observable process with measurement errors follows the multivariate functional autoregressive (MFAR) process. We later prove that such an observable process follows a multivariate functional autoregressive moving average (FARMA) process. Thus, existing approaches based on FPCA produce inefficient estimates.

This paper proposes a hierarchical model for multivariate functional time series, which is the extension of Kowal, Matteson \& Ruppert  \cite{kowal2019functional} to the multivariate setting. This model will overcome the issue of inconsistent solutions arising in the FPCA approach and produce efficient results in a sparse setting. To estimate the MFAR model, we firstly form a two-level hierarchical model using the observed values. For computation, we later represent the hierarchical model as a dynamic linear model (DLM). We use observation equations to discretize the functional data and obtain measurement error. The evolution equations take into account the correlation among the functional time series and define the process model. The latent process in our case is the multivariate functional autoregressive model of order $p$, written as MFAR($p$). A dynamic functional factor model further specifies the dynamic innovation process. We employ an efficient Gibbs sampling algorithm to perform inference and prediction for multivariate time series.

While studying the cross-correlation among the bivariate functional time series, a natural question is whether there exists a directionality of information. We are interested in analyzing the causal relations among a set of variables. As discussed by Granger \cite{granger1969investigating}, the first time series $X$ is said to have a causal influence on the second time series $Y$ if the prediction of second series $Y$ could be improved by the inclusion of past measurements from the first time series $X$. Later, classical Granger causality was extended to different types of time series model. Boudjellaba, Dufour, \& Roy \cite{boudjellaba1994simplified} derived the necessary and sufficient condition for non-causality in stationary vector autoregressive (VAR) process. Comte \& Lieberman \cite{comte2000second} investigated second-order non-causality in the framework of the VARMA process with generalized autoregressive conditional heteroscedasticity (GARCH) type errors. This unified treatment of first and second-order causality is widely used with some modification in various fields  (see, e.g., Hafner \& Herwartz \cite{hafner2008testing}, Wo{\'z}niak \cite{wozniak2015testing}, Chen, Marron \& Zhang \cite{chen2019modeling}). For most of these models, restrictions on parameters are linear functions. Droumaguet \& Wo{\'z}niak \cite{droumaguet2012bayesian} applied Markov-switching VAR models for which the restrictions for non-causality in variance are a non-linear function of parameters and used the Bayesian approach to test non-causality. Allen \& Hooper \cite{allen2018generalized} used Granger causality generalized measures of correlation (GcGMC) to analyze the causal relations between VIX and S\&P 500.

Above mentioned literature has applied some modified versions of Granger causality to multivariate time series. Saumard \cite{saumard2017linear} investigated the causality of functional time series by testing the equality of covariance operators for dependent processes. Recently, Shang, Ji \& Beyaztas \cite{shang2021granger} extended the GcGMC to bivariate functional time series. This paper will address the Granger causality for functional time series utilizing the Bayes factor technique introduced by Kass \& Raftery \cite{kass1995bayes}. Bayes factor provides a way of evaluating evidence in favor of the null hypothesis; in our case, it tests the non-causality. Following Geweke \cite{geweke1999using}, we use a sample of the posterior distribution to evaluate the model's marginal likelihood, estimate the integral over the posterior sample, and calculate the modified harmonic mean (MHM). Using the marginal density of data of the two models under consideration, we calculate and use the Bayes factor to choose the preferred model.

The rest of the paper is organized as follows. In Section 2 we illustrate the MFAR model of order $p$ and its representation in a dynamic linear model (DLM), which is further used for inference and prediction. We then generalize the concept of Granger causality to functional time series. Section 3 provides the methodology followed in this paper including a detailed explanation of the Bayes factor and the method of obtaining the same. Section 4 presents the empirical results of checking the causal relationship among the bond yields of different countries and effect of meteorological factors on air pollution level. Section 5 presents the concluding remarks.

\section{The Model}
Suppose we have $K$ functional time series defined as $\{Y_t^n\},\:n=1,\cdots,K$ on some compact index set $\mathcal{T}\subset \mathrm{R}^D$, typically $D=1$. Consider $\{\bm{Y}_t\}$ as $K-$variate functional time series represented as: $\bm{Y}_t = [Y_t^1,\cdots,Y_t^K]'$ where $Y_t^n \in \mathrm{L^2(\mathcal{T})}$ for $n = 1, 2, \cdots,K.$
\subsection{Multivariate Functional Autoregressive Model}
Multivariate functional autoregressive model of order $p$, written as MFAR($p$):
\begin{equation}\label{eq1}
  \bm{Y}_t - \bm{\mu} = \sum_{l=1}^p\bm{\Psi}_l(\bm{Y}_{t-l} - \bm{\mu})+\bm{\epsilon}_t.
\end{equation}
where $\bm{\mu} = [\mu^1,\mu^2,\cdots,\mu^K]'$ is mean of $\bm{Y}_t$ under stationarity, $\bm{\epsilon}_t = [\epsilon_t^1,\epsilon_t^2,\cdots, \epsilon_t^K]'$ is a vector of innovation functions with each $\epsilon_t^n$ being a sequence of independent mean zero random functions with $\mathrm{E}\mid \mid \epsilon_t^n\mid \mid ^2<\infty$ for $n=1,\cdots,K$ and
$\bm{\Psi}_l= [\Psi^{i,j}_l]_{i,j=1}^K$
where each $\Psi^{i,j}_l$ is a bounded linear operator on $\mathrm{L^2(\mathcal{T})}$ for $i,j = 1,\cdots,K$ and $l=1,2,\cdots,p$.

We assume that each function $Y_t^n(\tau^n),\:n=1,\cdots,K$ is observed at unequally-spaced points $\tau^n \in\mathcal{T}$. We restrict to $p=1$ and consider only integral operators for better interpretability and computational convenience. In practice, the functional observations $\{Y_t\}$ are observed via discrete samples of each curve with some measurement error. Suppose we observe $y_{i,t}^n \in \mathrm{R}$ sampled with measurement error $v_{i,t}^n$ from $\{Y_t^n\}\in \mathrm{L^2}(\mathcal{T}),$ $n=1,\cdots,K$:
\begin{equation}
     y_{i,t}^n = Y_t^n(\tau_{i,t}^n)+v_{i,t}^n
\end{equation}
where $\tau_{i,t}^n$ for $i=1,\cdots,m_t^n$ are observation points of $Y_t^n$ and $v_{i,t}^n$ is a mean zero measurement error with finite variance. Assume $\bm{\alpha}_t = \bm{Y}_t - \bm{\mu}$, then two level hierarchical model is given as:
\begin{equation}\label{eq5}
\begin{aligned}
    y_{i,t}^n  &= \mu^n(\tau^n_{i,t})+\alpha^n_t(\tau^n_{i,t}) +v^n_{i,t}, \hspace{32mm} i = 1,\cdots,m_t^n,\\
    \alpha_t^n &= \sum_{m=1}^K\int\psi^{n,m}(\tau^n,u)\alpha_{t-1}^m(u)du + \epsilon_t^n(\tau^n), \hspace{7mm}\forall\tau\in\mathcal{T} \text{ and } n=1,\cdots,K,
\end{aligned}
\end{equation}
for $t=2,\cdots, T,$ where $\psi^{n,m}$ are integral operators and we assume that $\{\bm{v}_{i,t}\}$ and $\{\bm{\epsilon}_t\}$ are mutually independent.

The presence of measurement error in model (\ref{eq5}) increases the estimation error of $\bm{\Psi}$ and produces inefficient forecasts. The following proposition illustrates that MFAR($p$) model for the observables is inappropriate.

\textbf{Proposition 1:} Let $\bm{Y}_t-\bm{\mu} = \sum_{l=1}^p \bm{\Psi}(\bm{Y}_{t-l}+\bm{\epsilon}_t)$ and suppose we observe $\bm{y}_t = \bm{Y}_t+\bm{v}_t$ are independent white noise processes. Then the observable process $\{\bm{y}_t\}$ follows a multivariate functional autoregressive moving average (FARMA) process of order ($p, p$).\\
For proof see Kowal et al. \cite{kowal2019functional}.

To overcome this model mis-specification, we separate the observed data into functional process and measurement errors in the dynamic linear model (DLM).
\subsection{Dynamic Linear Models for MFAR($p$)}
For practical implementation of model (\ref{eq5}), we select a finite set of evaluation points for each of functional time series, $\mathcal{T}_e^n = \{\tau_1^n,\cdots,\tau_M^n\} \subset\mathcal{T}$ for $n=1,\cdots,K.$ We approximate the integral in (\ref{eq5}) using quadrature methods
with $Q^{n,m}$ for $n,m = 1,\cdots,K,$ as a known quadrature weight matrix. Let $Z_t^n$ be $m_t^n\times M,\: n=1,\cdots,K,$ incidence matrix that identifies the observation points observed at time $t$ for $n$-th functional times series. We can formulate the dynamic linear model (DLM) corresponding to the hierarchical model (\ref{eq5}) as:
\begin{equation}\label{eq7}
    \begin{aligned}
    \bm{y}_t &= \bm{Z}_t\bm{\mu} +\bm{Z}_t\bm{\alpha}_t + \bm{v}_t, \hspace{10mm}
       [\bm{v}_t\mid \Sigma_v]\stackrel{indep}{\sim}\mathit{N}(\bm{0},\Sigma_v) \: \text{ for } t=1,\cdots,T,\\
    \bm{\alpha}_t &= \bm{\psi}\bm{Q}\bm{\alpha}_{t-1}+
    \bm{\epsilon}_t, \hspace{10mm} [\bm{\epsilon}_t\mid \bm{K}_{\epsilon}] \stackrel{indep}{\sim}\mathit{N}(\bm{0},
    \bm{K}_{\epsilon})\:\text{ for } t=2,\cdots,T,\\
    \bm{\alpha}_1 &\sim \mathit{N}(\bm{0},\bm{K}_{\epsilon})
    \end{aligned}
\end{equation}

where $\bm{y}_t = [y_{i,t}^n],$ is a discrete sample of $K$-variate functional time series $\{\bm{Y}_t\} \text{ for } i=1,\cdots,m_t^n \text{ and } n = 1,\cdots,K.$ Here, $\bm{\mu} = [\mu^1(\tau^1),\cdots, \mu^K(\tau^K)]'$ and  $\bm{\alpha}_t = [\alpha_{t}^1(\tau^1),\cdots,  \alpha_{t}^K(\tau^K)]'$ where $\mu^i(\tau^i)=[\mu^i(\tau^i_1),\cdots,\mu^i(\tau^i_M)]$ and $\alpha_{t}^i(\tau^i) = [\alpha_{t}^i(\tau^i_1),\cdots,\alpha_{t}^i(\tau^i_M)]$ for $i=1,\cdots,K.$ Other matrices are defined as:
\begin{equation*}
\bm{Z}_t = \begin{bmatrix}
Z_t^1 & 0&\cdots&0\\
\vdots& &\ddots&\vdots\\
0 & 0&\cdots& Z_t^K\\
\end{bmatrix},\:
\bm{\psi} = \begin{bmatrix}
\{\psi^{1,1}(\tau_i,\tau_j)\}_{i,j =1}^M&\cdots&\{\psi^{1,K}(\tau_i,\tau_j)\}_{i,j =1}^M\\
\vdots &\ddots&\vdots\\
\{\psi^{1,K}(\tau_i,\tau_j)\}_{i,j =1}^M&\cdots&\{\psi^{K,1}(\tau_i,\tau_j)\}_{i,j =1}^M\\
\end{bmatrix}\end{equation*}
\begin{equation*}\text{ and }
\bm{Q} = \begin{bmatrix}
Q^{1,1}&\cdots&Q^{1,K}\\
\vdots &\ddots&\vdots\\
Q^{1,K}&\cdots&Q^{K,1}
\end{bmatrix}.
\end{equation*}
Model (\ref{eq7}) can be extended to multiple lags to the MFAR($p$) by replacing the evolution equation with $ \bm{\alpha}_t = \sum_{l=1}^p\bm{\psi}_l\bm{Q}\bm{\alpha}_{t-l}+\bm{\epsilon}_t$.

\subsection{Granger Causality in Functional Autoregressive Model}
Here, we investigate the causal relationship between two functional time series data $Y_t$ and $X_t$, $t\in T$ observed on unequally spaced points $\tau \in \mathcal{T}$ where $\mathcal{T}\subset \mathbf{R}$ is a compact index set. Firstly, we model $Y_t$ as functional autoregressive model of order 1, written as FAR(1), and partition the linear projection of $Y_t$ on $Y_{t-1}$ and $X_{t-1}$ to account for the Granger causality as
\begin{MEquation}
\label{m1}
\begin{pmatrix}
     Y_t -\mu_Y \\
      X_t -\mu_X
\end{pmatrix} =
\begin{bmatrix}
\Psi_{YY} & \Psi_{YX}\\
0 & \Psi_{XX}
\end{bmatrix}
\begin{pmatrix}
     Y_{t-1}-\mu_Y \\
     X_{t-1}-\mu_X
\end{pmatrix}+\begin{pmatrix}
     \epsilon_{Y_t}\\
     \epsilon_{X_t}
\end{pmatrix}
\end{MEquation}
In second model, we model $Y_t$ as linear projection of $Y_{t-1}$ only as
\begin{MEquation}
\label{m2}
    \begin{pmatrix}
     Y_t -\mu_Y \\
      X_t -\mu_X
\end{pmatrix} =
\begin{bmatrix}
\Psi'_{YY} & 0\\
0 & \Psi'_{XX}
\end{bmatrix}
\begin{pmatrix}
     Y_{t-1}-\mu_Y \\
     X_{t-1}-\mu_X
\end{pmatrix}+\begin{pmatrix}
     \epsilon'_{Y_t}\\
     \epsilon'_{X_t}
\end{pmatrix}
\end{MEquation}
where $Y_t$ and $X_t \in \mathrm{L^2(\mathcal{T})}$ and $\mu_Y$ and $\mu_X$ are mean of $\{Y_t\}$ and $\{X_t\}$ respectively. Consider $\epsilon_t = (\epsilon_{Y_t} \:\: \epsilon_{X_t})'$ and $\epsilon'_t = (\epsilon'_{Y_t} \:\: \epsilon'_{X_t})'$ as zero mean independent random functions with covariance matrices $K_{\epsilon}$ and $K_{\epsilon'}$ respectively. In model (\ref{m1}), $Y_t$ depends on both $Y_{t-1}$ and $X_{t-1}$ and is considered as unrestricted model while model (\ref{m2}) is considered as restricted as $Y_t$ depends only it's past values. In this setting, $X$ is said to Granger cause $Y$ if model (\ref{m1}) is significantly better than model (\ref{m2}) at forecasting future values of $Y$.

Model (\ref{m1}) and (\ref{m2}) are in MFAR form, but for the computation purpose, we need to convert the MFAR models into DLM form as described in equation (\ref{eq7}). Let $\Theta_1$ and $\Theta_2$ be the vector of parameters corresponding to model (\ref{m1}) and (\ref{m2}), respectively. Now, $\Theta_1 = (\mu,\Sigma_{\nu},\Psi,K_{\epsilon})$ and $\Theta_2 = (\mu,\Sigma_{\nu},\Psi',K_{\epsilon'})$ where,
\begin{equation*}
    \mu = \begin{bmatrix}
    \mu_Y\\
    \mu_X
    \end{bmatrix},\:
    \Sigma_{\nu}= \begin{bmatrix}
    \sigma^2_{\nu_Y} & 0\\
    0 & \sigma^2_{\nu_X}
    \end{bmatrix},\:
    \Psi = \begin{bmatrix}
    \Psi_{YY} & \Psi_{YX}\\
0 & \Psi_{XX}
    \end{bmatrix}
   \text{ and }
    \Psi' = \begin{bmatrix}
\Psi'_{YY} & 0\\
0 & \Psi'_{XX}
\end{bmatrix}.
\end{equation*}

\section{Methodology} We hypothesize that the unrestricted model (\ref{m1}) yields better prediction in contrast to the restricted model (\ref{m2}). The null hypothesis indicates no  additional contribution from the past of $X$ to the future evolution of $Y$ once the past of $Y$ has been take into account. The alternative is that there is an additional contribution. In the univariate setting the usual way to test for Granger causality is to look at the forecast error sum of squares from each mode land compare them through an F test. For multivariate data, this can be extended by taking the trace or determinant of the covariance matrices under each model. As the dimension of the data becomes large, the $F$ test becomes more unreliable as the estimation of the covariance matrix in higher dimensions is not efficient. For functional data, which is infinite dimensional, this procedure is unusable. Hence we look into Bayesian procedures which concentrate on the dimension of the parameters and not of the data.

The Bayes factor is a central quantity of interest in Bayesian hypothesis testing. The ratio of the marginal data densities for both hypothesis-specific models is known as the Bayes factor. Let $p(Y_t\mid M)$ be the marginal data density for model $M$; then the Bayes factor would be

\begin{equation}\label{bayes}
    B_{12} = \frac{p(Y_t\mid \ref{m1})}{p(Y_t\mid \ref{m2})}.
\end{equation}

Bayesian hypothesis testing begins by specifying different prior distributions on the parameters involved in the models. We also have the likelihood for the data given the prior values of the parameter. Using these, the crucial step is the computation of marginal data densities, $p(Y_t\mid M)$ for both models.  For computing this value, we apply the Modified Harmonic Mean (MHM) method. Let $\Theta$ be a K-variate vector of parameters corresponding to the model.  Firstly, obtain the sample of draws $\{\Theta^{(i)}\}_{i=1}^S$, where $S$ is the number of samples, using the posterior distribution of the parameters $p(\Theta\mid Y_t)$ for each model. Now, the marginal density of data is computed as:
\begin{equation}\label{marginaldensity}
    p(Y_t\mid M)= \left[S^{-1}\sum_{i=1}^S\frac{h(\Theta^{(i)})}{\mathcal{L}(Y_t;\Theta^{(i)})p(\Theta^{(i)})}\right]^{-1}
\end{equation}
where, $\mathcal{L}(Y_t;\Theta^{(i)})$ is the likelihood function and $p(\Theta^{(i)})$ is the prior density function of parameters obtained by substituting sample of draws $\{\Theta^{(i)}\}_{i=1}^S$. Also, $h(\Theta^{(i)})$ is a K-variate truncated Normal distribution with mean equal to posterior mean of $\Theta$ and variance set to the posterior covariance matrix of $\Theta$.

In the rest of this section we elaborate the various steps involved in this procedure. In sections \ref{sec:prior} and \ref{sec:likelihood} we present the assumed prior distribution and the likelihood of the data under the two models. In section \ref{sec:posterior} we describe the method to draw samples from the posterior distribution of the parameter. Finally in section \ref{sec:algo} we present the algorithm to compute the Bayes factor.

\subsection{Prior Distribution}\label{sec:prior}
We specify the prior distributions for each of parameter in the vector $\Theta = (\mu,\Sigma_{\nu},\Psi,\epsilon_t)$ below.
\begin{enumerate}
    \item[1.] Prior distribution for mean function, $\mu$:
    \begin{align*}
    \mu(\tau) & = b'_{\mu}(\tau)\theta_{\mu}, \hspace{15mm} \text{ for }\tau \in \mathcal{T}, \:\: \mathcal{T} \text{ is the index set,}\\
    \text{where, } b'_{\mu} & \text{ is low rank thin plate spline basis and }\\
    \theta_{\mu} & \sim \mathit{N}(0, \Lambda_{\mu})\\
    s.t.\:\: & \Lambda_{\mu} = diag(10^8,10^8,\lambda_{\mu}^{-1},\lambda_{\mu}^{-1},\cdots,\lambda_{\mu}^{-1})\:\text{  and  }\: \lambda_{\mu}^{-1/2} \sim \text{Uniform}(0,10^4).
    \end{align*}
    \item[2.] Prior distribution for observed covariance matrix, $\Sigma_{\nu}$:
    \begin{align*}
        \Sigma_{\nu} & = \begin{bmatrix}
    \sigma^2_{\nu_Y} & 0\\
    0 & \sigma^2_{\nu_X}
    \end{bmatrix} \hspace{85mm}\\
    where,\:\: & \sigma^2_{\nu_Y} \sim \text{Gamma}(10^{-3},10^{-3})\:\text{ and }\:\sigma^2_{\nu_X} \sim \text{Gamma}(10^{-3},10^{-3}).
    \end{align*}
    \item[3.] Prior distribution for FAR kernel, $\Psi$:
    \begin{align*}
    \Psi(\tau,u) & = (b'_{\Psi}(u)\otimes b'_{\Psi}(\tau))\theta_{\Psi}, \hspace{10mm} \text{ for }\tau,u \in \mathcal{T},\\
    \text{where, } & b'_{\Psi} \text{ is cubic B-spline basis and }\\
    & [\theta_{\Psi}\mid \lambda_{\Psi}]  \sim \mathit{N}(0, \lambda_{\Psi}^{-1}\Omega_{\Psi}^{-1}) \hspace{10mm} (\text{$\theta_{\Psi}$ induces a Gaussian prior on $\Psi$})  \\
    s.t.\:\: & \lambda_{\Psi} = \Tilde{\zeta}_{\Psi}^{-2}\Tilde{\lambda}_{\Psi}, \hspace{10mm} \text{where, } \Tilde{\zeta}_{\Psi} \sim \mathit{N}(0,10^6) \:\text{ and }\:\Tilde{\lambda}_{\Psi} \sim \text{Gamma}(1/2,1/2) \:\text{ and }\\
    & \Omega_{\Psi} = \Omega_2 + \kappa \Omega_0, \hspace{5mm} \text{where, } \log(\kappa)\sim \mathit{N}(0,4), \:\: \Omega_0 = \int\int \Psi(\tau,u)d\tau du  \:\text{ and }\\
    &\hspace{10mm} \Omega_2 = \int\int\left(\frac{d^2}{d\tau^2}\Psi(\tau,u) + \frac{d^2}{d\tau du}\Psi(\tau,u)+\frac{d^2}{du^2} \Psi(\tau,u)\right) d\tau du.
    \end{align*}
    \item[4.] Prior distribution for evolution error, $\epsilon_t$:\\
    We use the functional dynamic linear model (FDLM) approach of Kowal et.al. (2016), in which they decomposes the evolution error $\epsilon_t$ into factor loading curves $\phi_j$, and time-dependent factors $e_{j,t} \in \mathbf{R}$, for $j=1,\cdots,J_{\epsilon}$:
    \begin{equation}
    \label{eq8}
        \epsilon_t(\tau) = \sum_{j=1}^{J_\epsilon}e_{j,t}\phi_j(\tau)+\eta_t(\tau), \hspace{10mm}\forall \tau\in\mathcal{T},
    \end{equation}
    where, $J_\epsilon$ is the number of factors and $\{\eta_t\}$ is mean zero approximation error. Priors for each of the parameters in equation (\ref{eq8}) are as follows:
    \begin{enumerate}
        \item Prior distribution for $\{\eta_t\}$:
        \begin{align*}
            \eta_t & \stackrel{iid}{\sim} \mathcal{GP}(0,K_\eta)\hspace{5mm}
            \text{where, }\:K_\eta(\tau,u)=\sigma_\eta^2\bm{1}(\tau=u)\\
            &\hspace{27mm} s.t.\:\: \sigma_\eta^{-2}\sim Gamma(10^{-3},10^{-3}) \text{ and } \bm{1}(.) \text{ is the indicator function.}
        \end{align*}
        \item Prior distribution for $e_t$:
        \begin{align*}
            & e_t\sim\mathit{N}(0,\Sigma_e) \hspace{10mm}
            \text{where, }\:\Sigma_e = diag(\{\sigma_j^2\}_{j=1}^{J_\epsilon}).\\
           \text{Joint distribution of }& [\sigma_1^{-2},\cdots,\sigma_{J_\epsilon}^{-2}] =[\sigma_{J_\epsilon}^{-2}]\prod_{j=1}^{J_\epsilon -1} [\sigma_j^{-2}\mid \sigma_{j+1}^{-2},\cdots,\sigma_{J_\epsilon}^{-2}]\\
           \text{where, }  \sigma_{J_\epsilon}^{-2}  \sim  \text{Gamma} &(10^{-3},10^{-3}) \:\:\:\text{ and}\\
           [\sigma_j^{-2}\mid  \sigma_{j+1}^{-2},\cdots,\sigma_{J_\epsilon}^{-2}] & = [\sigma_j^{-2}\mid \sigma_{j+1}^{-2}] \sim \text{Uniform}(0,\sigma_{j+1}^{-2})\hspace{5mm}\text{for } j=1,\cdots,J_\epsilon-1.
        \end{align*}
        \item Prior distribution for $\phi_j$ for $j=1,\cdots,J_\epsilon$:
        \begin{align*}
        \phi_j(\tau) & = b'_{\phi}(\tau)\theta_{\phi}, \hspace{15mm} \text{ for }\tau \in \mathcal{T},\\
        \text{where, } b'_{\phi} & \text{ is low rank thin plate spline basis and }\\
        \theta_{\phi} & \sim \mathit{N}(0, \Lambda_{\phi})\\
        s.t.\:\: & \Lambda_{\phi} = diag(10^8,10^8,\lambda_{\phi}^{-1},\lambda_{\phi}^{-1},\cdots,\lambda_{\phi}^{-1})\:\text{  and  }\: \lambda_{\phi}^{-1/2} \sim \text{Uniform}(0,10^4).
        \end{align*}
    \end{enumerate}
\end{enumerate}

\subsection{Likelihood Function}\label{sec:likelihood}
We use the dynamic linear model described in equation ($\ref{eq7}$) to form the likelihood function for $Y_t$. At initial time $i=1$, we have
\begin{equation*}
  \begin{aligned}
  \bm{Y}_1 &\sim N(\bm{Z}_1 \bm{\mu}^{(1)} + \bm{Z}_1 \bm{\alpha}_1, \Sigma_v^{(1)})\\
  \bm{\alpha}_1 &\sim N(\bm{0},\bm{K_{\epsilon}}^{(1)})
  \end{aligned}
\end{equation*}
Now, for $i = 2,\cdots,n$, one-step ahead distribution of $Y_t$ is given as:
\begin{equation*}
    \begin{aligned}
     \bm{Y}_{i\mid i-1} &\sim N(\bm{Z}_i \bm{\mu}^{(i)} + \bm{Z}_i \bm{\alpha}_{i\mid i-1}, \Sigma_v^{(i)}  + \bm{Z}_i\bm{\Xi}_{i\mid i-1}\bm{Z}_i^T )\\
     \bm{\alpha}_{i\mid i-1} &= \bm{\psi}^{(i)}\bm{\alpha}_{i-1\mid i-1}, \hspace{5mm} \text{and}\hspace{5mm} \bm{\Xi}_{i\mid i-1} = \bm{\psi}^{(i)}\bm{\Xi}_{i-1\mid i-1} [\bm{\psi}^{(i)}]^T + \bm{K_{\epsilon}}^{(i)})
    \end{aligned}
\end{equation*}
Priors for each time is given as:
\begin{equation*}
    \begin{aligned}
     \bm{\alpha}_{i\mid i} &= \bm{\alpha}_{i\mid i-1} + \bm{K}_i \bm{v}_i, \hspace{5mm} \text{and}\hspace{5mm}
     \bm{\Xi}_{i\mid i} = (I_M - \bm{K}_i\bm{Z}_i)\bm{\Xi}_{i\mid i-1},\\
     \text{where},\:\:\: K_i &= \bm{\Xi}_{i\mid i-1} \bm{Z}_i^T(\bm{Z}_i\bm{\Xi}_{i\mid i-1}\bm{Z}_i^T + \Sigma_v)^{-1},\hspace{5mm} \text{and}\hspace{5mm}
     v_i = (\bm{Y}_{i\mid i-1} - \bm{Z}_i \bm{\mu}^{(i)}) -  \bm{Z}_i \bm{\alpha}_{i\mid i-1}.\\
    \end{aligned}
\end{equation*}
\subsection{Posterior Sampling Distribution}\label{sec:posterior}
We use Gibbs sampling algorithm to obtain the posterior sampling distribution. Firstly, we initialize the DLM (\ref{eq7}) parameters and later we present the sampling distributions.
\subsubsection*{Initialization}
\begin{itemize}
    \item[Step 1:] We firstly initialize $\mu = [\mu_Y\:\: \mu_X]'$ as smooth mean of $[\{y_t^n\}\:\: \{x_t^n\}]_{t=1}^{'T}$ and fit a spline to each $[Y_t - \mu_Y\:\: X_t - \mu_X]'$ to estimate $\bm{\alpha}_t$. Thus, we obtain the estimates of $\mathrm{E}[Y_t\:\: X_t]'$, which is further used in estimating the measurement error variance-covariance matrix $\Sigma_{\bm{v}}$:
    \begin{equation*}
        \Sigma_{\bm{v}} = \begin{bmatrix}
        \sigma^{-2}_{v_Y}&0\\
       0&\sigma^{-2}_{v_X}\\
       \end{bmatrix},
    \end{equation*}
where $\sigma^{-2}_{v_Y} = \sum_{t=1}^T (y_t - \mathrm{E}[Y_t])$ and $\sigma^{-2}_{v_X} = \sum_{t=1}^T (x_t - \mathrm{E}[X_t])$.

\item[Step 2:] Now, we use $\bm{\alpha}_t$ to initialize the FAR kernels $\bm{\psi}$. For $\bm{\psi} = B_{\bm{\psi}} \theta_{\bm{\psi}}$
, where $B_{\bm{\psi}}=(b_{\bm{\psi}}(\tau_1),\cdots,b_{\bm{\psi}}(\tau_M))$ for $M$ number of evaluation points and $\theta_{\bm{\psi}}$ is sampled from $[\theta_{\bm{\psi}}\mid \cdots] \sim N(A_{\bm{\psi}} a_{\bm{\psi}}, A_{\bm{\psi}}),$ where
\begin{equation*}
\begin{aligned}
A_{\bm{\psi}}^{-1}  &=
\lambda_{\bm{\psi}}\Omega_{\bm{\psi}} + [B_{\bm{\psi}}'\{\sum_{t=p+1}^T \bm{\alpha}'_{t-1}\bm{\alpha}_{t-1} \}B_{\bm{\psi}}] [B_{\bm{\psi}}'B_{\bm{\psi}}], \\
a_{\bm{\psi}} &= vec\left(B_{\bm{\psi}}' \{\sum_{t=p+1}^T \bm{\alpha}'_{t-1}\bm{\alpha}_{t-1}\} B_{\bm{\psi}}'\right)\\
\end{aligned}
\end{equation*}
where, $B_{\bm{\psi}}$ is a cubic B-spline basis functions with equally spaced knots, $\Omega_{\bm{\psi}}$ is a prior precision matrix and $\lambda_{\bm{\psi}} = 1$, is a smoothing parameter.

\item[Step 3:] Using estimates for FAR kernels in second equation of model (\ref{eq7}), we compute the innovations $\bm{\epsilon}_t$. We decompose the innovations into factor loading curves (FLCs) $\Phi = (\phi_1,\cdots,\phi_{J_{\bm{\epsilon}}})$, and time-dependent factors, $e_t = (e_{1,t},\cdots,e_{J_\varepsilon,t})'$ as described in equation (\ref{eq8}). For simplicity, we assume $[e_t\mid\Sigma_e] \stackrel{indep}{\sim} N(0,\Sigma_e),$ with $\Sigma_e = diag(\{\sigma^2_j\}_{j=1}^{J_{\bm{\epsilon}}}),$ where $J_{\bm{\epsilon}}$ is the number of factors. We initialize the FDLM parameters using the singular value decomposition (SVD) of $(\bm{\epsilon}_1, \cdots, \bm{\epsilon}_T)' = U_eD_eV_e'.$ We set $\Phi$ equal to the first $J_{\bm{\epsilon}}$ columns of $V_e$ and the factors $(e_1, \cdots,e_T)'$ equal to the first $J_{\bm{\epsilon}}$ columns of $(U_eD_e)$. Further, we estimate $\{\sigma^2_j\}$ and $\sigma^2_{\eta}$ using the conditional maximum likelihood estimators and $\{\eta_t\}$ is the mean zero approximation error with $\eta_t \stackrel{iid}{\sim} GP(0, \sigma_{\eta}^2).$

\item[Step 4:] The implied covariance matrix for $\bm{\epsilon}_t$ is $K_{\bm{\epsilon}} = \Phi\Sigma_e\Phi'+\sigma^2_{\eta}I_M$, conditional on ${\phi_j,\sigma^2_j}$ and $\sigma^2_{\eta}$. We obtain the inverse of $K_{\bm{\epsilon}}$ using the Woodbury identity:
$$K_{\bm{\epsilon}}^{-1} = \sigma^{-2}_{\eta}I_M - \sigma^{-2}_{\eta} \Phi\Tilde{\Sigma_e}\Phi',$$
where $\Tilde{\Sigma_e}= diag(\{\sigma^2_j/(\sigma^2_{\eta}+\sigma^2_j)\}_{j=1}^{J_{\bm{\epsilon}}}).$
\end{itemize}

\subsubsection*{Sampling Distribution}
\begin{itemize}
    \item[Step 1:] We sample the FAR kernels $\bm{\psi}$, $\bm{\psi}_ = B_{\bm{\psi}}\theta_{\bm{\psi}},$ , using parametrization of  $\theta_{\bm{\psi}} = \Tilde{\xi}_{\bm{\psi}}\Tilde{\theta}_{\bm{\psi}}$, where we sample $\Tilde{\theta}_{\bm{\psi}}$ from $[\Tilde{\theta}_{\bm{\psi}}\mid\cdots]\sim N(A_{\bm{\psi}}a_{\bm{\psi}},A_{\bm{\psi}}),$ where
    \begin{equation*}
\begin{aligned}
A_{\bm{\psi}}^{-1}  &=
\lambda_{\bm{\psi}}\Omega_{\bm{\psi}} +  s_l \Tilde{\xi}_{\bm{\psi}_l}^2[(B_{\bm{\psi}}'\bm{Q})\{\sum_{t=p+1}^T \bm{\alpha}'_{t-1}\bm{\alpha}_{t-1} \}(B_{\bm{\psi}}'\bm{Q})']\otimes [B_{\bm{\psi}}'K_{\bm{\epsilon}}^{-1}B_{\bm{\psi}}],\\
a_{\bm{\psi}} &=  \Tilde{\xi}_{\bm{\psi}} vec\left(B_{\bm{\psi}}' \{\sum_{t=p+1}^T \bm{\alpha}'_{t-1}\bm{\alpha}_{t-1}\} (B_{\bm{\psi}}'\bm{Q})'\right).
\end{aligned}
\end{equation*}
Sample $[\Tilde{\xi}_{\bm{\psi}}\mid\cdots]\sim N(A_{\Tilde{\xi}_{\bm{\psi}}}a_{\Tilde{\xi}_{\bm{\psi}}},A_{\Tilde{\xi}_{\bm{\psi}}}),$ where
\begin{equation*}
    \begin{aligned}
    A_{\Tilde{\xi}_{\bm{\psi}}}^{-1}  & = 10^{-6} + \Tilde{\theta}_{\bm{\psi}}'\left(\left[(B_{\bm{\psi}}'\bm{Q})\left(\sum_{t=p+1}^T \bm{\alpha}'_{t-1}\bm{\alpha}_{t-1} \right)(B_{\bm{\psi}}'\bm{Q})'\right] \otimes \left[B_{\bm{\psi}}'K_{\bm{\epsilon}}^{-1}B_{\bm{\psi}}\right]\right)\Tilde{\theta}_{\bm{\psi}}\\
    a_{\bm{\psi}} &= \Tilde{\theta}_{\bm{\psi}}' vec\left(B_{\bm{\psi}}' \{\sum_{t=p+1}^T \left[\bm{\alpha_t}-\sum_{k\neq 1} s_k \bm{\alpha}'_{t-k}\right]\bm{\alpha}_{t-1}\} (B_{\bm{\psi}}'\bm{Q})'\right)
    \end{aligned}
\end{equation*}
and sample $[\lambda_{\bm{\psi}}\mid\cdots]\sim Gamma\left(1/2+J_{\bm{\psi}}^2/2, 1/2+ \theta_{\bm{\psi}}'\Omega_{\bm{\psi}}\theta_{\bm{\psi}}/2\right)$.

\item[Step 2:] We compute the innovations using the estimate of FAR kernels. Under FDLM, we decompose the innovations into FLCs and time-dependent factors.
\begin{itemize}
    \item[(a)] The factors $\{e_t\}_{t=1}^T$: sample $[e_t\mid\cdots]\sim N(A_e a_{e_t},A_e)$, where
    $$A_e^{-1}=diag\left(\{\sigma^2_{\eta}+\sigma^2_j)\}_{j=1}^{J_{\bm{\epsilon}}}\right),$$
    $$a_{e_t} = \sigma^{-2}_{\eta}\Phi'\bm{\epsilon}_t.$$
    \item[(b)] The factor precision, $\sigma_j^{-2}$: sample
       \begin{equation*}
           [\sigma_{J_{\bm{\epsilon}}}^{-2}\mid\cdots]\sim Gamma\left(10^{-3}+\frac{T}{2}, 10^{-3}+ \frac{1}{2}\sum_{t=1}^T e_{J_{\bm{\epsilon}},t}^2\right).
       \end{equation*}
    \item[(c)] The approximation error precision, $\sigma_{\eta}^{-2}$: sample
    $$[\sigma_{\eta}^{-2}\mid\cdots]\sim Gamma\left(10^{-3}+\frac{TM}{2}, 10^{-3}+ \frac{1}{2}\sum_{t=1}^T \mid\mid\bm{\epsilon}_t - \Phi e_t\mid\mid^2\right),$$
    where $\mid\mid\cdot\mid\mid^2$ denotes the Euclidean distance and $M$ is the number of evaluation points .
    \item[(d)] The factor loading curves $\Phi = B_{\phi}\Xi$, where $B_{\phi}$ is the matrix of basis functions and $\Xi=(\xi_1,\cdots,\xi_{J_{\bm{\epsilon}}})$. For $j=1,\cdots,J_{\bm{\epsilon}}$, sample $\xi_j \sim N(A_{\xi_j}a_{\xi_j},A_{\xi_j})$, where
    $$A_{\xi_j}^{-1} = \sigma_{\eta}^{-2}\left(\sum_{t=1}^T e_{j,t}^2\right)B_{\phi}'B{\phi},$$
    $$a_{\xi_j} = \sigma_{\eta}^{-2}B_{\phi}'\sum_{t=1}^T e_{j,t}\left(\bm{\epsilon}_t - B_{\phi}\sum_{k\neq j}\xi_k e_{k,t}\right).$$
\end{itemize}
\item[Step 3:] We obtain the covariance matrix $K_{\bm{\epsilon}}$ and its inverse $K_{\bm{\epsilon}}^{-1}$ using the similar approach as in Step 4 of initialization.
\item[Step 4:] We sample measurement error precision, $\sigma_{\nu_Y}^{-2}$ and $\sigma_{\nu_X}^{-2}$ as follows:
$$[\sigma_{\nu_Y}^{-2}\mid\cdots] \sim Gamma\left(10^{-3}+\frac{1}{2}\sum_{t=1}^T m_t, 10^{-3}+\frac{1}{2}\sum_{t=1}^T(y_t-\mu_Y-\alpha_t)^2\right),$$
$$[\sigma_{\nu_X}^{-2}\mid\cdots] \sim Gamma\left(10^{-3}+\frac{1}{2}\sum_{t=1}^T m_t, 10^{-3}+\frac{1}{2}\sum_{t=1}^T(x_t-\mu_X-\alpha_t)^2\right).$$
\item[Step 5:] Form the DLM and sample $[\bm{\alpha}_{t=1}^T\mid\cdots],$ using KFAS package in R. We sample the mean function  $\mu = B_{\phi}'\theta_{\mu}$, where $[\theta_{\mu}\mid\cdots]\sim N(A_{\mu}a_{\mu},A_{\mu})$ where
$$A_{\mu}^{-1} = \Lambda_{\mu}^{-1}+\sigma_{\nu}^{-2}\sum_{t=1}^T B_{\phi}' (Z_t)' Z_t B_{\phi},$$
$$a_{\mu}=\sigma_{\nu}^{-2}\sum_{t=1}^T B_{\phi}'(Z_t)'(w_t -Z_t\alpha_t),$$
where, $ w_t = [y_t \:\: x_t]', \: \Lambda_{\mu} = diag(10^8,10^8,\lambda_{\mu}^{-1},\cdots,\lambda_{\mu}^{-1})$ and $[\lambda_{\mu}\mid\cdots]\sim Gamma\left(\frac{1}{2}(J_{\mu}-3),\frac{1}{2}\sum_{j=3}^{J_{\mu}}\theta_{\mu,j}^2 \right)$.
\end{itemize}

\subsection{Algorithm and Interpretation of Bayes Factor}\label{sec:algo}
\subsubsection*{Algorithm to compute Bayes factor:}
\begin{enumerate}
    \item Translate the models \ref{m1} and \ref{m2} into DLM form as described in equation (\ref{eq7}).
    \item Consider model \ref{m1}. Initialize the parameter space $\Theta_1 = (\mu,\Sigma_{\nu},\Psi,\epsilon_t)$ using the steps mentioned in Section \ref{sec:posterior} under Initialization.
    \item For $i = 1,\cdots,S$, where, $S$ is the number of simulations:
    \begin{enumerate}
        \item Obtain the sample of $\{\Theta_1^{(i)}\}$ by applying the posterior sampling distribution given in Section \ref{sec:posterior} under Sampling.
        \item Substitute the posterior sample $\{\Theta_1^{(i)}\}$ in prior distributions mentioned in Section \ref{sec:prior} to obtain prior density function $p(\Theta_1^{(i)})$.
        \item Compute the likelihood function $\mathcal{L}(Y_t;\Theta_1^{(i)})$ as described in Section \ref{sec:likelihood} and obtain a truncated Normal distribution with mean equal to posterior mean of $\Theta_1$ and variance set to the posterior covariance matrix of $\Theta_1$, denoted by $h(\Theta_1^{(i)})$.
        \item Substitute the values obtained from prior density function $p(\Theta_1^{(i)})$, likelihood function $\mathcal{L}(Y_t;\Theta_1^{(i)})$ and truncated Normal density function $h(\Theta_1^{(i)})$ in equation (\ref{marginaldensity}) to compute the marginal data density corresponding to model \ref{m1}, $p(Y_t\mid\ref{m1})$.
    \end{enumerate}
    \item Repeat step 2 and 3, considering model \ref{m2} and parameter space as $\Theta_2 = (\mu,\Sigma_{\nu},\Psi',\epsilon'_t)$ to obtain the marginal data density corresponding to model \ref{m2}, $p(Y_t\mid\ref{m2})$.
    \item Compute Bayes factor $B_{12}$ using the marginal data densities of model \ref{m1} and \ref{m2} and equation (\ref{bayes}).
\end{enumerate}

\subsubsection*{Interpretation of Bayes factor:}
Kass \& Raftery \cite{kass1995bayes} provided the interpretation of the Bayes factor values in terms of weighing evidence against the null hypothesis, presented in Table \ref{interpretation}. A positive logarithm of the Bayes factor is interpreted as evidence in favor of the unrestricted model. Symmetrically, a negative logarithm of the Bayes factor indicates that the data prefer the restricted model.
\begin{table}
\begin{center}
    \caption{Interpretation of Bayes Factor}
    \label{interpretation}
    \begin{tabular}{| l| l| l| }
        \hline\noalign{\smallskip}
    $ln(B_{12})$ & $B_{12}$ & Evidence against $H_0$ \\
      \noalign{\smallskip}\hline\noalign{\smallskip}
      0 to 1 & 1 to 3  & Not worth more than a bare mention \\
      1 to 3 & 3 to 20 & Substantial\\
      3 to 5 & 20 to 150 & Strong \\
      $>$5 & $>$150 & Decisive\\
      	    \noalign{\smallskip}\hline
    \end{tabular}
\end{center}
\end{table}

\section{Simulation study}
In this section,  we compared the predicted results of simulated functional time series using the proposed method for MFAR relative to FAR. We are particularly interested in one-step and five-step forecasting and finding the causal relationship between two functional time series. Throughout the experiments, we consider two scenarios based on model (\ref{m1}) and model (\ref{m2}) with varying sample size $T$. The FAR(1) kernel used for model (\ref{m2}) is the Bimodal-Gaussian kernel, $\Psi(\tau, u) \propto \frac{0.75}{\pi(0.3)(0.4)}\exp{−(\tau − 0.2)^2/(0.3)^2 − (u− 0.3)^2/(0.4)^2} +\frac{0.45}{\pi(0.3)(0.4)}\exp{−(\tau − 0.7)^2/(0.3)^2 − (u − 0.8)^2/(0.4)^2}$ and MFAR(1) kernels used for model (\ref{m1}) are also Bimodal-Gaussian kernal, $\Psi(\tau, u) = \Psi_Y(\tau, u)$ and $\Psi_X(\tau, u) \propto \frac{0.75}{\pi(0.35)(0.35)}\exp{−(\tau − 0.5)^2/(0.35)^2 − (u− 0.3)^2/(0.35)^2} +\frac{0.45}{\pi(0.35)(0.35)}\exp{−(\tau − 0.75)^2/(0.35)^2 −  (u − 0.6)^2/(0.35)^2}$. For stationarity, each kernel is rescaled to pre-specified squared norm, $C_{\Phi} = \int\int\Psi^2(\tau,u)d\tau du$, with $C_{\Phi}<1$. For FAR(1) we select $C_{\Psi} = 0.5$ and for MFAR(1) we select $C_{\Psi_Y},C_{\Psi_X} = (0.5,0.1)$. Thus, functional time series $\{Y_t\}$ generated from MFAR(1) kernels is correlated to functional time series $\{X_t\}$ while $\{Y_t\}$ generated from FAR(1) kernel is independent of $\{X_t\}$. We use smooth Gaussian process for innovation process $\epsilon_t$. In our experiment, we consider dense sampling design using $m_t = 30$ equally-spaced observation points on $[0, 1]$ for all $t$, however the model works well with sparse sampling design also. Data sets generated using FAR(1) kernel and MFAR(1) kernels are applied to proposed model (MFAR) as well as to univariate model proposed by Kowal et. al. (FAR) \cite{kowal2019functional}. Table \ref{RMSE} shows the one- and five-step root mean squared forecasting errors (RMSFEs) for both the models. When $\{Y_t\}$ is dependent on $\{X_t\}$,we obtain lower root mean squared forecasting errors for the MFAR model than the FAR model suggesting MFAR performs better than FAR in forecasting the functional time series.

To acknowledge the causal relationship between $\{X_t\}$ and $\{Y_t\}$, we calculate the Bayes factor as described in section \ref{sec:algo}. We consider three different simulated cases: (1) $\{Y_t\}$ is generated from MFAR and $\{X_t\}$ is generated from FAR, and we check whether the $\{X_t\}$ series causes $\{Y_t\}$ series, (2) $\{X_t\}$ is generated from MFAR and $\{Y_t\}$ is generated from FAR, and we check for causality in the opposite direction, and (3) $\{X_t\}$ and $\{Y_t\}$ both are generated from FAR. Table \ref{Table_Bayes} presents the Bayes factor $B_{12}$ and logarithm of Bayes factor $ln(B_{12})$ values. Based on $ln(B_{12})$, we conclude that there is statistical evidence in favor of a causal relationship in first two cases and no such relationship exists when $\{X_t\}$ and $\{Y_t\}$ are independent.

\begin{table}
\centering
\caption{Simulation Results showing h-step ahead RMSFEs}
    \begin{tabular}{|l|ll||ll|}
    \hline
    & h=1 &    & h=5 & \\
    \hline
    Data Generation & RMSFE  & RMSFE  & RMSFE & RMSFE \\
     & (MFAR) & (FAR) & (MFAR) & (FAR) \\
    \hline
     $\{Y_t\}$ is dependent on $\{X_t\}$ &  0.009160474
 & 0.009312481  &  0.01064169 & 0.0110255 \\
    $\{Y_t\}$ is independent of $\{X_t\}$ &  0.009157562 & 0.00964527  & 0.01069097 & 0.008994967  \\
    \hline
    \end{tabular}

    \label{RMSE}
\end{table}

\begin{figure}
    \centering
    \includegraphics[width=0.8\textwidth]{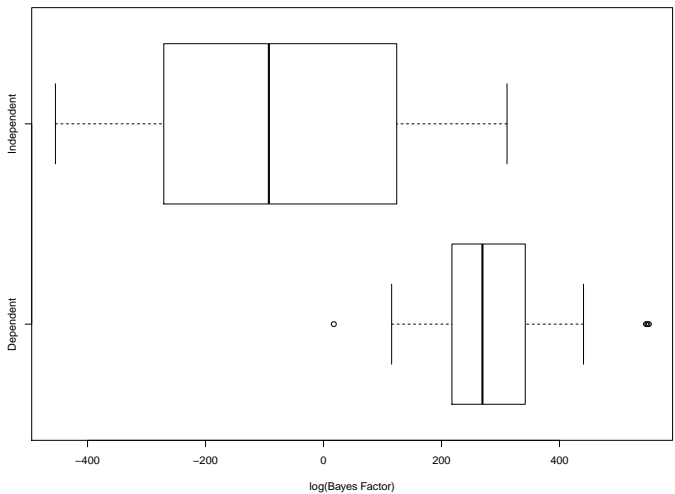}
    \caption{Boxplot of the logarithm of the Bayes factor for the dependent and independent cases.}
    \label{fig:boxplot}
\end{figure}

\begin{table}
\begin{center}
    \caption{Bayes Factor Values of Simulated Series}
    \label{Table_Bayes}
    \begin{tabular}{|l| l| l|l|}
        \hline
     & $\{X_t\}$ Cause $\{Y_t\}$ & $\{Y_t\}$ cause $\{X_t\}$ & $\{Y_t\}$ and $\{X_t\}$ are independent \\
      \hline
      $B_{12}$ & 8.70756E+33
 & 1.35857E+52 & 0.007122615\\
  $ ln(B_{12})$ & 78.14949918
 & 120.0408611 & -4.94448041\\
      	  \hline
    \end{tabular}
\end{center}
\end{table}
We now repeat the analysis presented in Table \ref{Table_Bayes} forty times for each of the dependent and independent cases and compute the Bayes factor. The boxplot of the logarithm of these Bayes factors is presented fin Figure \ref{fig:boxplot}. It can be seen that the distributions are well-separated. Therefore the method is useful in detecting whether the data arises from the dependent or the independent model.

\section{Application to Real Data}
We apply the proposed method to two real datasets. The first application is to daily data of yield curves over different maturities for two countries. For the second application, the response variable is still functional but the predictor variable is multivariate. The response is the daily curve of particulate matter in a city while the predictors are several meteorological variables.
\subsection{Testing causality between yield curves of two countries}
To assess the performance of the proposed model, we examine the causal relationship between the yield curves of two countries: USA and UK. The yield curve depicts the relationships of bond yields with different maturities. Yield curves have a significant impact on the money supply within the economy and thus give insights into the country's economic condition. Time series of yield curves have been studied by several authors. See Sen \& Kl{\"u}ppelberg \cite{sen2019time} for an introduction and exploratory plots. For our study, we considered 5-month data starting from Jan 2021 till May 2021, which includes 82 working days data. Values of the yield curve are available for 9 maturities: 3, 6, 12, 24, 36, 60, 84, 120, 360 months. We obtain the USA bond yield data for T maturities from the U.S. Department of the Treasury \href{website}{https://www.treasury.gov/resource-center/data-chart-center/interest-rates/Pages/TextView.aspx?data=realyield} and the UK bond yield data from \href{investing.com}{http://uk.investing.com}. The objective is to investigate whether UK yield curve Granger cause USA yield curve and vice-versa. We denote the USA yield curves as $Y_t$ and the UK yield curve as $X_t$, substituting them in models (\ref{m1}) and (\ref{m2}). Once we have the unrestricted and restricted models, we apply the Bayesian analysis to calculate the Bayes factor $B_{12}$ as defined in the equation (\ref{bayes}). We perform 1000 simulations to generate posterior samples of $\Theta$ of dimension 123. Looking at the logarithm of Bayes factor, $ln(B_{12})$ in table \ref{Case1}, we conclude that it is not statistically significant to say that the UK yield curve Granger cause USA yield curve. We also check whether the reverse relation exists or not, i.e., whether USA yield curve cause UK yield curve. In this case, we take  UK yield curves as $Y_t$ and the USA yield curve as $X_t$ in model (\ref{m1}) and (\ref{m2}). Based on $ln(B_{12})$ (obtained from table \ref{Case1}), we conclude that USA yield curve does not Granger cause UK yield curve. Now we reverse the predictor and response variables and repeat the analysis. We observe that the Bayes Factor is low in this case also and hence conclude that the UK yield curve does not Granger cause the USA yield curve.

\begin{table}
\begin{center}

    \caption{Bayes Factor Values of Yield Curves Example}
    \label{Case1}
    \begin{tabular}{l| l| l}
        \hline\noalign{\smallskip}
     & UK Cause USA & USA cause UK  \\
      \noalign{\smallskip}\hline\noalign{\smallskip}
     Bayes Factor $ (B_{12})$ & 3.18731E-20
 & 1.06315E-16 \\
 Logarithm(Bayes Factor) $ ln(B_{12})$ & -44.89252354 & -36.78012961 \\
      	    \noalign{\smallskip}\hline
    \end{tabular}
\end{center}
\end{table}

\subsection{Testing whether meteorological factors cause air pollution level}
\begin{figure}
    \centering
    \includegraphics[scale = 0.5]{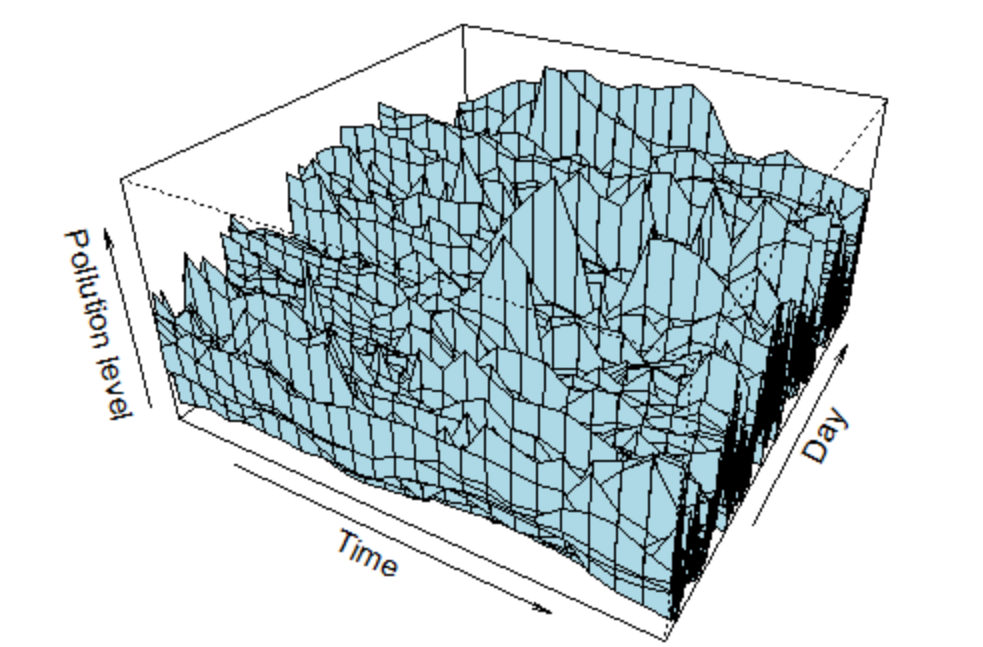}
    \caption{ The surface plot displays the evolution of the Pollution level in Delhi between the years
2015 and 2020, observed at every hour of a day.}
    \label{fig:pollution}
\end{figure}

Numerous pieces of literature have been forecasting air quality using meteorological features (see, e.g., Aneja et al. \cite{aneja2001measurements}, Kumar \& Goyal \cite{kumar2011forecasting}, Meng et al.\cite{meng2019contribution}). Here, we investigate the causal effect of meteorological factors such as temperature, humidity, wind speed, and solar radiation on air pollution concentration. To compute particle pollution levels, we use particulate matter with diameters 10 micrometers and smaller, denoted by PM10. For our study, we used daily data from Jan 2015 to June 2020 in Delhi, India, consisting of 2008 days. However, after data cleansing we are left with 1670 days of data. We downloaded the data from official Central Pollution Control Board of India \href{website}{https://cpcb.nic.in/}. We take $Y_t$ as PM10 in Delhi for each hour of a day; that is, [0,23] is the support of each function. Figure \ref{fig:pollution} provides with the surface
plot depicting the pollution level in Delhi over the years 2015 and 2020. For $X$, we take the daily average of three meteorological factors: temperature, relative humidity, and wind speed obtained from \href{meteoblue.com}{https://www.meteoblue.com/en/weather/week/new-delhi_india_1261481}. In this case, $X$ is a vector instead of a function. Using $X_t$ and $Y_t$, we obtain unrestricted and restricted models and apply the algorithm to compute the Bayes factor as mentioned in section \ref{sec:algo}. In this case, the estimated parameter space $\Theta$ is of length 361. The logarithm of the Bayes factor is obtained as $ln(B_{12}) = 78.61206652
$. Based on the interpretation of $ln(B_{12})$ from table (\ref{bayes}), we conclude that there is strong evidence that meteorological factors Granger cause air pollution.
\section{Conclusion}
The paper introduces the two-level hierarchical Gaussian process for modeling time series of multivariate functional data. We
examine the Granger causality among the functional responses utilizing Bayesian techniques. We compare the marginal data densities of the restricted and unrestricted model to form a conclusion on the causal relationship among the functional time series. We implement the procedure on two real datasets. For the data on yield curves we find that the yield curve of USA is not Granger caused by that of UK. Nor is the yield curve of UK Granger caused by that of USA. In the pollution example we find that daily pollution level curves are indeed Granger caused by meteorological factors. The details of simulations and applications to actual data can be accessed through the GitHub repository  \href{Bayesian-Testing-Of-Granger-Causality-In-Functional-Time-Series}{https://github.com/Shubhanghi/Bayesian-Testing-Of-Granger-Causality-In-Functional-Time-Series/branches}.

\bmhead{Acknowledgments} The research was supported by SERB-MATRICS grant MTR/2020/000324.

\bibliography{bib1212}


\begin{thebibliography}{33}
\ifx \bisbn   \undefined \def \bisbn  #1{ISBN #1}\fi
\ifx \binits  \undefined \def \binits#1{#1}\fi
\ifx \bauthor  \undefined \def \bauthor#1{#1}\fi
\ifx \batitle  \undefined \def \batitle#1{#1}\fi
\ifx \bjtitle  \undefined \def \bjtitle#1{#1}\fi
\ifx \bvolume  \undefined \def \bvolume#1{\textbf{#1}}\fi
\ifx \byear  \undefined \def \byear#1{#1}\fi
\ifx \bissue  \undefined \def \bissue#1{#1}\fi
\ifx \bfpage  \undefined \def \bfpage#1{#1}\fi
\ifx \blpage  \undefined \def \blpage #1{#1}\fi
\ifx \burl  \undefined \def \burl#1{\textsf{#1}}\fi
\ifx \doiurl  \undefined \def \doiurl#1{\url{https://doi.org/#1}}\fi
\ifx \betal  \undefined \def \betal{\textit{et al.}}\fi
\ifx \binstitute  \undefined \def \binstitute#1{#1}\fi
\ifx \binstitutionaled  \undefined \def \binstitutionaled#1{#1}\fi
\ifx \bctitle  \undefined \def \bctitle#1{#1}\fi
\ifx \beditor  \undefined \def \beditor#1{#1}\fi
\ifx \bpublisher  \undefined \def \bpublisher#1{#1}\fi
\ifx \bbtitle  \undefined \def \bbtitle#1{#1}\fi
\ifx \bedition  \undefined \def \bedition#1{#1}\fi
\ifx \bseriesno  \undefined \def \bseriesno#1{#1}\fi
\ifx \blocation  \undefined \def \blocation#1{#1}\fi
\ifx \bsertitle  \undefined \def \bsertitle#1{#1}\fi
\ifx \bsnm \undefined \def \bsnm#1{#1}\fi
\ifx \bsuffix \undefined \def \bsuffix#1{#1}\fi
\ifx \bparticle \undefined \def \bparticle#1{#1}\fi
\ifx \barticle \undefined \def \barticle#1{#1}\fi
\bibcommenthead
\ifx \bconfdate \undefined \def \bconfdate #1{#1}\fi
\ifx \botherref \undefined \def \botherref #1{#1}\fi
\ifx \url \undefined \def \url#1{\textsf{#1}}\fi
\ifx \bchapter \undefined \def \bchapter#1{#1}\fi
\ifx \bbook \undefined \def \bbook#1{#1}\fi
\ifx \bcomment \undefined \def \bcomment#1{#1}\fi
\ifx \oauthor \undefined \def \oauthor#1{#1}\fi
\ifx \citeauthoryear \undefined \def \citeauthoryear#1{#1}\fi
\ifx \endbibitem  \undefined \def \endbibitem {}\fi
\ifx \bconflocation  \undefined \def \bconflocation#1{#1}\fi
\ifx \arxivurl  \undefined \def \arxivurl#1{\textsf{#1}}\fi
\csname PreBibitemsHook\endcsname

\bibitem{guyet2016long}
\begin{barticle}
\bauthor{\bsnm{Guyet}, \binits{T.}},
\bauthor{\bsnm{Nicolas}, \binits{H.}}:
\batitle{Long term analysis of time series of satellite images}.
\bjtitle{Pattern Recognition Letters}
\bvolume{70},
\bfpage{17}--\blpage{23}
(\byear{2016})
\end{barticle}
\endbibitem

\bibitem{stoehr2021detecting}
\begin{barticle}
\bauthor{\bsnm{Stoehr}, \binits{C.}},
\bauthor{\bsnm{Aston}, \binits{J.A.}},
\bauthor{\bsnm{Kirch}, \binits{C.}}:
\batitle{Detecting changes in the covariance structure of functional time
  series with application to fmri data}.
\bjtitle{Econometrics and Statistics}
\bvolume{18},
\bfpage{44}--\blpage{62}
(\byear{2021})
\end{barticle}
\endbibitem

\bibitem{sen2019time}
\begin{barticle}
\bauthor{\bsnm{Sen}, \binits{R.}},
\bauthor{\bsnm{Kl{\"u}ppelberg}, \binits{C.}}:
\batitle{Time series of functional data with application to yield curves}.
\bjtitle{Applied Stochastic Models in Business and Industry}
\bvolume{35}(\bissue{4}),
\bfpage{1028}--\blpage{1043}
(\byear{2019})
\end{barticle}
\endbibitem

\bibitem{horvath2014testing}
\begin{barticle}
\bauthor{\bsnm{Horv{\'a}th}, \binits{L.}},
\bauthor{\bsnm{Kokoszka}, \binits{P.}},
\bauthor{\bsnm{Rice}, \binits{G.}}:
\batitle{Testing stationarity of functional time series}.
\bjtitle{Journal of Econometrics}
\bvolume{179}(\bissue{1}),
\bfpage{66}--\blpage{82}
(\byear{2014})
\end{barticle}
\endbibitem

\bibitem{li2020long}
\begin{barticle}
\bauthor{\bsnm{Li}, \binits{D.}},
\bauthor{\bsnm{Robinson}, \binits{P.M.}},
\bauthor{\bsnm{Shang}, \binits{H.L.}}:
\batitle{Long-range dependent curve time series}.
\bjtitle{Journal of the American Statistical Association}
\bvolume{115}(\bissue{530}),
\bfpage{957}--\blpage{971}
(\byear{2020})
\end{barticle}
\endbibitem

\bibitem{chen2017adaptive}
\begin{barticle}
\bauthor{\bsnm{Chen}, \binits{Y.}},
\bauthor{\bsnm{Li}, \binits{B.}}:
\batitle{An adaptive functional autoregressive forecast model to predict
  electricity price curves}.
\bjtitle{Journal of Business \& Economic Statistics}
\bvolume{35}(\bissue{3}),
\bfpage{371}--\blpage{388}
(\byear{2017})
\end{barticle}
\endbibitem

\bibitem{chen2019modeling}
\begin{barticle}
\bauthor{\bsnm{Chen}, \binits{Y.}},
\bauthor{\bsnm{Marron}, \binits{J.}},
\bauthor{\bsnm{Zhang}, \binits{J.}}:
\batitle{Modeling seasonality and serial dependence of electricity price curves
  with warping functional autoregressive dynamics}.
\bjtitle{The Annals of Applied Statistics}
\bvolume{13}(\bissue{3}),
\bfpage{1590}--\blpage{1616}
(\byear{2019})
\end{barticle}
\endbibitem

\bibitem{aue2015prediction}
\begin{barticle}
\bauthor{\bsnm{Aue}, \binits{A.}},
\bauthor{\bsnm{Norinho}, \binits{D.D.}},
\bauthor{\bsnm{H{\"o}rmann}, \binits{S.}}:
\batitle{On the prediction of stationary functional time series}.
\bjtitle{Journal of the American Statistical Association}
\bvolume{110}(\bissue{509}),
\bfpage{378}--\blpage{392}
(\byear{2015})
\end{barticle}
\endbibitem

\bibitem{besse2000autoregressive}
\begin{barticle}
\bauthor{\bsnm{Besse}, \binits{P.C.}},
\bauthor{\bsnm{Cardot}, \binits{H.}},
\bauthor{\bsnm{Stephenson}, \binits{D.B.}}:
\batitle{Autoregressive forecasting of some functional climatic variations}.
\bjtitle{Scandinavian Journal of Statistics}
\bvolume{27}(\bissue{4}),
\bfpage{673}--\blpage{687}
(\byear{2000})
\end{barticle}
\endbibitem

\bibitem{ramsay1991some}
\begin{barticle}
\bauthor{\bsnm{Ramsay}, \binits{J.O.}},
\bauthor{\bsnm{Dalzell}, \binits{C.}}:
\batitle{Some tools for functional data analysis}.
\bjtitle{Journal of the Royal Statistical Society: Series B (Methodological)}
\bvolume{53}(\bissue{3}),
\bfpage{539}--\blpage{561}
(\byear{1991})
\end{barticle}
\endbibitem

\bibitem{bosq2000linear}
\begin{bbook}
\bauthor{\bsnm{Bosq}, \binits{D.}}:
\bbtitle{Linear Processes in Function Spaces: Theory and Applications}
vol. \bseriesno{149}.
\bpublisher{Springer},
\blocation{Germany}
(\byear{2000})
\end{bbook}
\endbibitem

\bibitem{ramsay2005principal}
\begin{bbook}
\bauthor{\bsnm{Ramsay}, \binits{J.}},
\bauthor{\bsnm{Silverman}, \binits{B.}}:
\bbtitle{Functional Data Analysis},
pp. \bfpage{147}--\blpage{172}.
\bpublisher{Springer},
\blocation{New York}
(\byear{2005})
\end{bbook}
\endbibitem

\bibitem{chiou2014multivariate}
\begin{botherref}
\oauthor{\bsnm{Chiou}, \binits{J.-M.}},
\oauthor{\bsnm{Chen}, \binits{Y.-T.}},
\oauthor{\bsnm{Yang}, \binits{Y.-F.}}:
Multivariate functional principal component analysis: A normalization approach.
Statistica Sinica,
1571--1596
(2014)
\end{botherref}
\endbibitem

\bibitem{zhu2012multivariate}
\begin{barticle}
\bauthor{\bsnm{Zhu}, \binits{H.}},
\bauthor{\bsnm{Li}, \binits{R.}},
\bauthor{\bsnm{Kong}, \binits{L.}}:
\batitle{Multivariate varying coefficient model for functional responses}.
\bjtitle{Annals of statistics}
\bvolume{40}(\bissue{5}),
\bfpage{2634}
(\byear{2012})
\end{barticle}
\endbibitem

\bibitem{li2017functional}
\begin{barticle}
\bauthor{\bsnm{Li}, \binits{J.}},
\bauthor{\bsnm{Huang}, \binits{C.}},
\bauthor{\bsnm{Hongtu}, \binits{Z.}},
\bauthor{\bsnm{Initiative}, \binits{A.D.N.}}:
\batitle{A functional varying-coefficient single-index model for functional
  response data}.
\bjtitle{Journal of the American Statistical Association}
\bvolume{112}(\bissue{519}),
\bfpage{1169}--\blpage{1181}
(\byear{2017})
\end{barticle}
\endbibitem

\bibitem{zhu2017multivariate}
\begin{barticle}
\bauthor{\bsnm{Zhu}, \binits{H.}},
\bauthor{\bsnm{Morris}, \binits{J.S.}},
\bauthor{\bsnm{Wei}, \binits{F.}},
\bauthor{\bsnm{Cox}, \binits{D.D.}}:
\batitle{Multivariate functional response regression, with application to
  fluorescence spectroscopy in a cervical pre-cancer study}.
\bjtitle{Computational statistics \& data analysis}
\bvolume{111},
\bfpage{88}--\blpage{101}
(\byear{2017})
\end{barticle}
\endbibitem

\bibitem{zhu2016bayesian}
\begin{barticle}
\bauthor{\bsnm{Zhu}, \binits{H.}},
\bauthor{\bsnm{Strawn}, \binits{N.}},
\bauthor{\bsnm{Dunson}, \binits{D.B.}}:
\batitle{Bayesian graphical models for multivariate functional data}.
\bjtitle{Journal of Machine Learning Research}
\bvolume{17},
\bfpage{1}--\blpage{27}
(\byear{2016})
\end{barticle}
\endbibitem

\bibitem{li2020fast}
\begin{barticle}
\bauthor{\bsnm{Li}, \binits{C.}},
\bauthor{\bsnm{Xiao}, \binits{L.}},
\bauthor{\bsnm{Luo}, \binits{S.}}:
\batitle{Fast covariance estimation for multivariate sparse functional data}.
\bjtitle{Stat}
\bvolume{9}(\bissue{1}),
\bfpage{245}
(\byear{2020})
\end{barticle}
\endbibitem

\bibitem{kowal2019functional}
\begin{barticle}
\bauthor{\bsnm{Kowal}, \binits{D.R.}},
\bauthor{\bsnm{Matteson}, \binits{D.S.}},
\bauthor{\bsnm{Ruppert}, \binits{D.}}:
\batitle{Functional autoregression for sparsely sampled data}.
\bjtitle{Journal of Business \& Economic Statistics}
\bvolume{37}(\bissue{1}),
\bfpage{97}--\blpage{109}
(\byear{2019})
\end{barticle}
\endbibitem

\bibitem{granger1969investigating}
\begin{barticle}
\bauthor{\bsnm{Granger}, \binits{C.W.}}:
\batitle{Investigating causal relations by econometric models and
  cross-spectral methods}.
\bjtitle{Econometrica: journal of the Econometric Society}
\bvolume{37}(\bissue{3}),
\bfpage{424}--\blpage{438}
(\byear{1969})
\end{barticle}
\endbibitem

\bibitem{boudjellaba1994simplified}
\begin{barticle}
\bauthor{\bsnm{Boudjellaba}, \binits{H.}},
\bauthor{\bsnm{Dufour}, \binits{J.-M.}},
\bauthor{\bsnm{Roy}, \binits{R.}}:
\batitle{Simplified conditions for noncausality between vectors in multivariate
  arma models}.
\bjtitle{Journal of Econometrics}
\bvolume{63}(\bissue{1}),
\bfpage{271}--\blpage{287}
(\byear{1994})
\end{barticle}
\endbibitem

\bibitem{comte2000second}
\begin{barticle}
\bauthor{\bsnm{Comte}, \binits{F.}},
\bauthor{\bsnm{Lieberman}, \binits{O.}}:
\batitle{Second-order noncausality in multivariate garch processes}.
\bjtitle{Journal of time series analysis}
\bvolume{21}(\bissue{5}),
\bfpage{535}--\blpage{557}
(\byear{2000})
\end{barticle}
\endbibitem

\bibitem{hafner2008testing}
\begin{barticle}
\bauthor{\bsnm{Hafner}, \binits{C.M.}},
\bauthor{\bsnm{Herwartz}, \binits{H.}}:
\batitle{Testing for causality in variance using multivariate garch models}.
\bjtitle{Annales d'Economie et de Statistique}
\bvolume{89},
\bfpage{215}--\blpage{241}
(\byear{2008})
\end{barticle}
\endbibitem

\bibitem{wozniak2015testing}
\begin{barticle}
\bauthor{\bsnm{Wo{\'z}niak}, \binits{T.}}:
\batitle{Testing causality between two vectors in multivariate garch models}.
\bjtitle{International Journal of Forecasting}
\bvolume{31}(\bissue{3}),
\bfpage{876}--\blpage{894}
(\byear{2015})
\end{barticle}
\endbibitem

\bibitem{droumaguet2012bayesian}
\begin{botherref}
\oauthor{\bsnm{Droumaguet}, \binits{M.}},
\oauthor{\bsnm{Wo{\'z}niak}, \binits{T.}}:
Bayesian testing of granger causality in markov-switching vars
(2012)
\end{botherref}
\endbibitem

\bibitem{allen2018generalized}
\begin{barticle}
\bauthor{\bsnm{Allen}, \binits{D.E.}},
\bauthor{\bsnm{Hooper}, \binits{V.}}:
\batitle{Generalized correlation measures of causality and forecasts of the vix
  using non-linear models}.
\bjtitle{Sustainability}
\bvolume{10}(\bissue{8}),
\bfpage{2695}
(\byear{2018})
\end{barticle}
\endbibitem

\bibitem{saumard2017linear}
\begin{bchapter}
\bauthor{\bsnm{Saumard}, \binits{M.}}:
\bctitle{Linear causality in the sense of granger with stationary functional
  time series}.
In: \bbtitle{Functional Statistics and Related Fields},
pp. \bfpage{225}--\blpage{231}.
\bpublisher{Springer},
\blocation{Switzerland}
(\byear{2017})
\end{bchapter}
\endbibitem

\bibitem{shang2021granger}
\begin{barticle}
\bauthor{\bsnm{Shang}, \binits{H.L.}},
\bauthor{\bsnm{Ji}, \binits{K.}},
\bauthor{\bsnm{Beyaztas}, \binits{U.}}:
\batitle{Granger causality of bivariate stationary curve time series}.
\bjtitle{Journal of Forecasting}
\bvolume{40}(\bissue{4}),
\bfpage{626}--\blpage{635}
(\byear{2021})
\end{barticle}
\endbibitem

\bibitem{kass1995bayes}
\begin{barticle}
\bauthor{\bsnm{Kass}, \binits{R.E.}},
\bauthor{\bsnm{Raftery}, \binits{A.E.}}:
\batitle{Bayes factors}.
\bjtitle{Journal of the american statistical association}
\bvolume{90}(\bissue{430}),
\bfpage{773}--\blpage{795}
(\byear{1995})
\end{barticle}
\endbibitem

\bibitem{geweke1999using}
\begin{barticle}
\bauthor{\bsnm{Geweke}, \binits{J.}}:
\batitle{Using simulation methods for bayesian econometric models: inference,
  development, and communication}.
\bjtitle{Econometric reviews}
\bvolume{18}(\bissue{1}),
\bfpage{1}--\blpage{73}
(\byear{1999})
\end{barticle}
\endbibitem

\bibitem{aneja2001measurements}
\begin{barticle}
\bauthor{\bsnm{Aneja}, \binits{V.P.}},
\bauthor{\bsnm{Agarwal}, \binits{A.}},
\bauthor{\bsnm{Roelle}, \binits{P.A.}},
\bauthor{\bsnm{Phillips}, \binits{S.B.}},
\bauthor{\bsnm{Tong}, \binits{Q.}},
\bauthor{\bsnm{Watkins}, \binits{N.}},
\bauthor{\bsnm{Yablonsky}, \binits{R.}}:
\batitle{Measurements and analysis of criteria pollutants in new delhi, india}.
\bjtitle{Environment International}
\bvolume{27}(\bissue{1}),
\bfpage{35}--\blpage{42}
(\byear{2001})
\end{barticle}
\endbibitem

\bibitem{kumar2011forecasting}
\begin{barticle}
\bauthor{\bsnm{Kumar}, \binits{A.}},
\bauthor{\bsnm{Goyal}, \binits{P.}}:
\batitle{Forecasting of air quality in delhi using principal component
  regression technique}.
\bjtitle{Atmospheric Pollution Research}
\bvolume{2}(\bissue{4}),
\bfpage{436}--\blpage{444}
(\byear{2011})
\end{barticle}
\endbibitem

\bibitem{meng2019contribution}
\begin{barticle}
\bauthor{\bsnm{Meng}, \binits{C.}},
\bauthor{\bsnm{Cheng}, \binits{T.}},
\bauthor{\bsnm{Gu}, \binits{X.}},
\bauthor{\bsnm{Shi}, \binits{S.}},
\bauthor{\bsnm{Wang}, \binits{W.}},
\bauthor{\bsnm{Wu}, \binits{Y.}},
\bauthor{\bsnm{Bao}, \binits{F.}}:
\batitle{Contribution of meteorological factors to particulate pollution during
  winters in beijing}.
\bjtitle{Science of The Total Environment}
\bvolume{656},
\bfpage{977}--\blpage{985}
(\byear{2019})
\end{barticle}
\endbibitem

\end{thebibliography}
\end{document}